\documentclass[reprint, twocolumn, superscriptaddress, preprintnumbers, amsmath, amssymb, aps, prc]{revtex4-2}

\allowdisplaybreaks
\allowdisplaybreaks[4]
\usepackage{graphicx}
\usepackage{dcolumn}
\usepackage{bm}
\usepackage{booktabs} 
\usepackage[
            pdfstartview=FitH,
            CJKbookmarks=true,
            bookmarksnumbered=true,
            bookmarksopen=true,
            colorlinks,
            linkcolor=blue,
            anchorcolor=blue,
            citecolor=blue
            ]{hyperref}
\usepackage{anyfontsize}
\usepackage{braket}
\usepackage{verbatim}
 \usepackage{multirow} 
 \usepackage{tabularx}
\newcommand\dd{\,\mathrm{d}}  %
\usepackage{mathtools}  %
\begin{document}

\title{Relativistic dynamical effects in proton emission: the Wentzel–Kramers–Brillouin method for 1+1 dimensional Dirac equation}

\author{Guangping Chen}
\affiliation{School of Physics and Astronomy, Beijing Normal University,
Beijing 100875, China}

\author{Wenmin Deng}
\affiliation{School of Physics and Astronomy, Beijing Normal University,
Beijing 100875, China}

\author{Ganlong Ding}
\affiliation{School of Physics and Astronomy, Beijing Normal University,
Beijing 100875, China}

\author{Sibo Wang}
\email[Corresponding author:~]{sbwang@cqu.edu.cn}
\affiliation{Department of Physics and Chongqing Key Laboratory for Strongly Coupled Physics, Chongqing University, Chongqing 401331, China}
\affiliation{Department of Physics, Graduate School of Science, The University of Tokyo, Tokyo 113-0033, Japan}

\author{Jing Peng}
\email[Corresponding author:~]{jpeng@bnu.edu.cn}
\affiliation{School of Physics and Astronomy, Beijing Normal University, Beijing 100875, China}
\affiliation{Key Laboratory of Multiscale Spin Physics (Ministry of Education), Beijing Normal University, Beijing 100875, China}

\author{Haozhao Liang}
\email[Corresponding author:~]{haozhao.liang@phys.s.u-tokyo.ac.jp}
\affiliation{Department of Physics, Graduate School of Science, The University of Tokyo, Tokyo 113-0033, Japan}
\affiliation{Quark Nuclear Science Institute, The University of Tokyo, Tokyo 113-0033, Japan}
\affiliation{RIKEN Center for Interdisciplinary Theoretical and Mathematical Sciences (iTHEMS), Wako 351-0198, Japan}

\date{\today}

\begin{abstract}

Starting from the $1+1$ dimensional (one spatial and one temporal dimension) Dirac equation, we employ the Wentzel–Kramers–Brillouin (WKB) approximation to derive the corresponding relativistic penetration probability. The derivation shows that the semiclassical momentum is determined by the Schr\"odinger-equivalent potential
\(
U_{\text{eff}}(r) = S(r) + \frac{E}{m}V(r) + \frac{S^{2}(r)-V^{2}(r)}{2m},
\)
instead of the simple sum of scalar and vector potentials \(S(r)+V(r)\), which has been adopted widely in the studies of relativistic quantum tunneling. We then quantify the relativistic dynamical effects in proton emission by comparing the results obtained with \(U_{\text{eff}}(r)\) and those obtained with \(S(r)+V(r)\). Incorporating \(U_{\text{eff}}(r)\) systematically reduces the penetration probability and the assault frequency, and consequently increases the predicted half-life. The relativistic dynamical effect becomes more pronounced with higher orbital angular momentum and can reach about \(84\%\) in the half-life of \(^{144}\mathrm{Tm}\).

\end{abstract}

\maketitle


\section{Introduction}\label{sec1}
Proton radioactivity, the spontaneous emission of protons from atomic nuclei, is a rare and intriguing mode of radioactive decay that occurs in proton-rich nuclei beyond the proton drip line~\cite{Qi2019214}. It was first experimentally confirmed in 1970 by Jackson \textit{et al.}~\cite{JACKSON1970281} and Cerny \textit{et al.}~\cite{CERNY1970284}, where the proton emission from the isomeric state of $^{53}$Co to the ground state of $^{52}$Fe was observed. Subsequently, ground-state proton radioactivity was discovered in $^{151}$Lu by Hofmann \textit{et al.}~\cite{Hofmann1982} and in $^{147}$Tm by Klepper \textit{et al.}~\cite{Klepper1982}. The study of proton emission plays a crucial role in validating nuclear models~\cite{Zhang2010, Delion2022, Zhai2024, FanYin2025}, extracting valuable spectroscopic information~\cite{Wang201783, Zhang2018}, and investigating nuclear deformation and structure~\cite{KARNY2008, Qi2012, Qian2013}.

In practical half-life calculations of proton emission, one of the most sensitive quantities is the penetration probability. Since the emission process is physically characterized as quantum tunneling through a sufficiently wide and high barrier, the Wentzel–Kramers–Brillouin (WKB) approximation~\cite{Basu200511, Dong200905, 2011-Sahu-Phys.Rev.C, Zhao201411, 2016-QianYB-Eur.Phys.J.A, Sahoo201904, Zhu202611} is highly applicable, with the decay width primarily determined by its characteristic semiclassical factor. Notably, this barrier-penetration picture is also widely extended to other processes, such as $\alpha$ decay ~\cite{Ni200810, Chandran202302, Luo202302} and cluster radioactivity~\cite{Ward201810, Mathson201904, Wang202104}. 
More generally, the WKB approximation is a standard asymptotic method for solving wave equations in situations where exact solutions are unavailable or analytically complicated~\cite{Landau1958, Heading1962, Froman1965, Griffiths_Schroeter_2018}. The evaluation of tunneling probabilities is one of its most important applications.

Once the WKB framework is adopted, the crucial ingredient is the potential barrier experienced by the emitted proton. In relativistic descriptions of nuclear structure, particularly within the relativistic mean-field (RMF) theory~\cite{1996-Ring-Prog.Part.Nucl.Phys., 2011-Niksic-Prog.Part.Nucl.Phys., 2015-LiangHZ-Phys.Rep.}, the nucleon dynamics is governed by the Dirac equation with a scalar potential $S(r)$ and a vector potential $V(r)$. In several recent practical proton-emission calculations based on RMF theory, the nuclear part of the barrier is taken as a simple form of $S(r)+V(r)$~\cite{Xia201601, yang2023, Lu2024, 2025-ZengLX-arXiv}. This choice is intuitively appealing, since it is closely associated with the dominant potential acting on the large component of the Dirac spinor.

However, the WKB penetration formula itself is originally derived from the Schr\"odinger equation. From the viewpoint of theoretical consistency, the quantity entering the semiclassical exponent should therefore not be simply $S+V$, but the Schr\"odinger-equivalent potential obtained from the Dirac equation~\cite{1980-Jaminon-PhysRevC.22.2027},
\begin{equation}
    U_{\text{eff}}(r) = S(r) + \frac{E}{m}V(r)+\frac{S^2(r)-V^2(r)}{2m},
\end{equation}
where $E$ is the single-particle energy including the proton mass $m$. Such a Schr\"odinger-equivalent potential has long been used in relativistic optical-model analyses~\cite{1981-Arnold-PhysRevC.23.1949, 1990-Hama-PhysRevC.41.2737, 2016-XuRR-PhysRevC.94.034606, 2024-QinPP-PhysRevC.109.064603}. Recent work has also shown that replacing the conventional choice $S+V$ by $U_{\text{eff}}$ from the relativistic microscopic optical potential~\cite{2024-QinPP-PhysRevC.109.064603, 2025-Qin-Nucl.Sci.Tech} may lead to a novel angular-dependent emission mechanism for proton emission~\cite{2026-FanYin-Phys.Rev.C}. It is therefore natural to ask what the essential form of the nuclear potential should be in a WKB treatment based on the Dirac equation. To answer this, it is necessary to solve the Dirac equation directly via the WKB approximation.

The WKB approximation applied to the Dirac equation has been discussed in several different contexts. For example, it has been used to analyze the Dirac equation in a supercritical Coulomb field~\cite{POPOV197868}, where the wave functions in the classically allowed and forbidden regions, together with their connection at the turning point, can be obtained. It has also been employed in Dirac equations with spherically symmetric scalar and vector potentials in quark systems, making it possible to study bound states and decay widths of hydrogen-like quark systems~\cite{Rubish2004}, as well as scaling properties related to quark-hadron duality~\cite{Orden200509}. These works demonstrate that combining the Dirac equation with the WKB approximation is inherently a well-established approach. Nevertheless, the question of which potential should consistently enter the relativistic WKB treatment of proton emission has not been clarified in a sufficiently clear way.

In this paper, we address this issue by deriving the WKB solution of the 1+1 dimensional (one spatial and one temporal dimension) Dirac equation. We demonstrate that the relevant quantity is $U_{\text{eff}}$ rather than the simple combination $S+V$. This paper is organized as follows. In Sec.~\ref{sec2}, we develop the theoretical framework for the penetration probability by solving the Dirac equation with the WKB approximation. In Sec.~\ref{sec4}, the numerical details are displayed. In Sec.~\ref{sec5}, we compare the penetration probabilities, assault frequencies, and half-lives obtained with the two choices of the nuclear potential, thereby quantifying the relativistic dynamical effects associated with replacing $S(r)+V(r)$ by $U_{\text{eff}}(r)$. In Sec.~\ref{sec6}, a summary is given.

\section{Theoretical framework}\label{sec2}

\subsection{WKB formalism for 1+1 dimensional Dirac equation}\label{secIIA}

As a starting point, we consider the 1+1 dimensional Dirac equation~\cite{Smirnov:2016efu, Greiner2003},
\begin{equation}
    \left\{-i\hbar\alpha\frac{\mathrm{d}}{\mathrm{d}x}+\beta\left[m+S(x)\right]+V(x)\right\}\Psi(x)=E\Psi(x),
    \label{eq-Dirac1+1}
\end{equation}
where $S(x)$ and $V(x)$ denote the scalar and vector potentials, respectively, and $E$ is the single-particle energy of a proton with mass $m$. Units with $c=1$ are used throughout this work. Choosing the representation of the Dirac gamma matrices
\begin{equation}
    \alpha=\begin{pmatrix}0&1\\ 1&0\end{pmatrix},\quad
    \beta=\begin{pmatrix}1&0\\ 0&-1\end{pmatrix},
    \label{eq-alphabeta}
\end{equation}
and decomposing the spinor $\Psi(x)$ as
\begin{equation}
    \Psi(x)=\begin{pmatrix}
      F(x)\\G(x)
    \end{pmatrix},
\end{equation}
we can reduce the Dirac equation~\eqref{eq-Dirac1+1} to the following coupled first-order equations:
\begin{subequations}
\begin{align}
    \frac{\mathrm{d}F}{\mathrm{d}x} +  \frac{1}{i\hbar} 
            \left[ \left( E - V \right) + \left( m + S \right) \right] G &= 0,\label{eq-dF-G}\\
    \frac{\mathrm{d}G}{\mathrm{d}x} + \frac{1}{i\hbar} 
            \left[ \left( E - V \right) - \left( m + S \right) \right] F &= 0.\label{eq-dG-F}
\end{align}
\end{subequations}
Eliminating the lower component $G(x)$ yields the following second-order differential equation for the upper component $F(x)$:
\begin{equation}\label{eq-ddF}
    \begin{split}
        \frac{\dd^2 F}{\dd x^2} - &\frac{S^\prime-V^\prime}{E-V+m+S}\frac{\dd F}{\dd x} \\
        \qquad \quad  +&\frac{1}{\hbar^2}\left[(E-V)^2-(m+S)^2\right]F=0.
    \end{split}
\end{equation}
Hereafter, the prime denotes differentiation with respect to $x$. To derive a Schr\"odinger-like equation suitable for the WKB approximation, it is necessary to eliminate the first-order derivative term in Eq.~\eqref{eq-ddF}. This can be accomplished by introducing the transformation~\cite{1985-Clark-book}
\begin{equation} \label{eq-Fxi}
    F(x)=\mathcal{A}^{1/2}(x)\chi(x),
\end{equation}
where
\begin{equation}\label{eq-Adef}
    \mathcal{A}(x)\equiv E-V(x)+m+S(x).
\end{equation}
This leads to
\begin{equation}
\begin{split}
   &\frac{\dd^2 \chi}{\dd x^2}+\left\{ \frac{1}{\hbar^2}\left[(E-V)^2-(m+S)^2\right] \right.  \\
   &\qquad \qquad + \left.\frac{S^{\prime\prime}-V^{\prime\prime}}{2\mathcal{A}(x)}-\frac{3(S^{\prime}-V^{\prime})^2}{4\mathcal{A}^2(x)}\right\}\chi=0.
    \label{eq-xi}   
\end{split}
\end{equation}
Equation~\eqref{eq-xi} takes a Schr\"odinger-like form, thus serving as the natural starting point for the WKB analysis in the relativistic framework.

We now apply the standard WKB ansatz~\cite{Dunham1932},
\begin{equation}\label{eq-chieta}
    \chi=e^{i\eta/\hbar},
\end{equation}
where the phase function $\eta(x)$ is expanded in powers of $\hbar$ as
\begin{equation}\label{eq-eta}
    \eta=\eta_{0}+\frac{\hbar}{i}\eta_{1}+\left(\frac{\hbar}{i}\right)^{2}\eta_{2}+\cdots
\end{equation}
Substituting Eqs.~\eqref{eq-chieta} and \eqref{eq-eta} into Eq.~\eqref{eq-xi} and collecting terms order by order in $\hbar$, we obtain a hierarchy of equations. The equations of the first three orders read
\begin{subequations}\label{eq-hbar123}
\begin{align}
    {\eta_{0}^{\prime}}^{2} &= (E-V)^2-(m+S)^2, \label{eq-hbar123-1}\\
    \eta_{0}^{\prime\prime}+2\eta_{0}^{\prime}\eta_{1}^{\prime} &= 0, \label{eq-hbar123-2} \\
    {\eta_{1}^{\prime}}^{2} + 2\eta_{0}^{\prime}\eta_{2}^{\prime} + \eta_{1}^{\prime\prime}
    &= \frac{1}{2}\frac{V^{\prime\prime}-S^{\prime\prime}}{\mathcal A(x)}
    + \frac{3}{4}\left(\frac{V^\prime-S^\prime}{\mathcal A(x)}\right)^2.
\end{align}
\end{subequations}
From the leading-order equation \eqref{eq-hbar123-1}, the zeroth-order phase is obtained as
\begin{equation}
    \eta_{0}(x)=\pm \int^{x}p(\xi)\dd{\xi },
\end{equation}
where the classical momentum $p(x)$ is defined by
\begin{equation}
    p(x)\equiv\sqrt{(E-V)^2-(m+S)^2}.
\end{equation}
By introducing the effective energy and effective potential as
\begin{subequations}\label{eq-eff-def}
\begin{align}
      E_{\text{eff}}&\equiv\frac{E^2-m^2}{2m}, \label{eq-Eeff}\\
      U_{\text{eff}}(x)&\equiv S(x) + \frac{E}{m}V(x)+\frac{S^2(x)-V^2(x)}{2m}, \label{eq-Ueff}
\end{align}
\end{subequations}
we can express the relativistic momentum in the nonrelativistic form
\begin{equation} \label{eq-momentum}
      p(x)=\sqrt{2m\left[ E_{\text{eff}}-U_{\text{eff}}(x) \right]}. 
\end{equation}
Equation~\eqref{eq-momentum} shows that the effective potential entering the relativistic WKB momentum is $U_{\text{eff}}(x)$ rather than the simple combination $S(x)+V(x)$. For proton emission, the single-particle energy is taken as $E = m+Q_p$, where $Q_p$ is the decay energy; consequently, the factor $E/m$ is typically close to unity. 
Nevertheless, the term $(S^2-V^2)/2m$ is not fundamentally negligible and may lead to a nontrivial deviation of $U_{\text{eff}}(x)$ from $S(x)+V(x)$. The quantitative impact of this difference on proton-emission half-lives will be examined in Sec.~\ref{sec5}.

The first-order correction $\eta_1(x)$ follows from Eq.~\eqref{eq-hbar123-2},
\begin{equation}
\begin{split}
    \eta_{1}(x)
        =& \int ^{x}\eta_{1}^{\prime}(\xi)\dd \xi
            =-\frac{1}{2}\int^{x} \frac{\eta_{0}^{\prime \prime}(\xi)}{\eta_{0}^{\prime}(\xi)}\dd \xi \\
        =& \ln p^{-1/2}(x)+C_0,
\end{split}
\end{equation}
where $C_0$ is an integration constant. Retaining terms up to $\mathcal{O}(\hbar)$ in Eq.~\eqref{eq-eta} yields the first-order WKB solution~\cite{Landau1958, Griffiths_Schroeter_2018}
\begin{equation}\label{eq-FWKB}
    F(x) = C\left[\frac{\mathcal{A}(x)}{p(x)}\right]^{1/2} \exp{\left(\pm \frac{i}{\hbar}\int^{x} p(\xi)\mathrm{d}\xi \right)},
\end{equation}
where the integration constant has been absorbed into the overall coefficient $C$. By substituting the WKB form of $F(x)$ into Eq.~\eqref{eq-dF-G}, we can express the lower component as
\begin{equation}
    G(x)=-\frac{i\hbar}{\mathcal{A}(x)}\frac{\dd F(x)}{\dd x} 
        =  \pm \frac{p(x)}{\mathcal{A}(x)} F(x) + \mathcal{O}(\hbar).
\end{equation}
Therefore, in the first-order WKB approximation, the Dirac spinor takes the form 
\begin{equation} \label{eq-psi-fg}
    \Psi(x)= 
    \begin{pmatrix}
      F(x)\\ \pm\frac{p(x)}{\mathcal{A}(x)}F(x)
    \end{pmatrix}.
\end{equation}

The first-order WKB approximation is valid provided that the quasi-classical condition
\begin{equation}
    \hbar |\eta_{0}^{\prime\prime}| \ll |\eta_{0}^{\prime 2}| \quad \text{or, equivalently,} \quad 
    \left|\frac{\mathrm{d}}{\mathrm{d}x}\left(\frac{\lambda}{2\pi}\right)\right| \ll 1,
    \label{eq-wkbcond}
\end{equation}
is satisfied, where the de Broglie wavelength $\lambda$ is given by
\begin{equation}
    \lambda=\frac{2\pi\hbar}{p}=\frac{2\pi\hbar}{\sqrt{2m(E_{\text{eff}}-U_{\text{eff}})}}.
    \label{eq-lam}
\end{equation} 

This condition means that the local wavelength varies slowly with position~\cite{Landau1958}. Consequently, the first-order WKB expressions derived above remain valid only in the regions where Eq.~\eqref{eq-wkbcond} holds. In particular, the approximation breaks down in the vicinity of the classical turning points where $p(x)=0$ (or equivalently, $E_{\text{eff}}=U_{\text{eff}}(x)$), causing the WKB wavefunction~\eqref{eq-FWKB} to become singular. As a result, the WKB solutions in the classically allowed and forbidden regions cannot be directly continued across the turning points merely through their local expressions. Instead, to accurately describe tunneling through the barrier, one must establish appropriate connection formulas between these regions, from which the final transmission probability can be derived.

\subsection{Connection formulas and penetration probability}\label{sec-IIB}

\begin{figure}[!ht]
    \centering
	\includegraphics[width=0.85\linewidth]{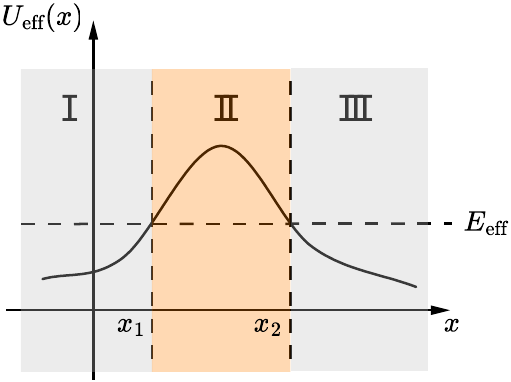}
	\caption{Effective potential barrier $U_{\mathrm{eff}}(x)$ with turning points $x_1$ and $x_2$ defined by $E_{\mathrm{eff}}=U_{\mathrm{eff}}(x)$. The grey-shaded regions I and III are classically allowed, while the orange-shaded region II is classically forbidden.}
	\label{fig1}
\end{figure}

We now turn to the tunneling process through the effective potential barrier $U_{\text{eff}}(x)$ depicted in Fig.~\ref{fig1}. The classical turning points $x_1$ and $x_2$ are determined by the condition
\begin{equation}
    E_{\text{eff}}=U_{\text{eff}}(x_1)=U_{\text{eff}}(x_2).
\end{equation} 
By applying the appropriate connection formulas to match the boundaries, we can unify the regional coefficients into a single overall coefficient $C_3$, expressing the upper component $F(x)$ across the three regions as
\begin{widetext}
\begin{equation}
		F(x)=
        \begin{cases}
		\kern-6pt & 2C_3e^{-\frac{i\pi}{4}}e^{\gamma}\sqrt{\frac{\mathcal{A}(x)}{p(x)}}\cos\left(\frac{1}{\hbar}\int^{x}_{x_1} p(\xi) \mathrm{d}\xi +\frac{\pi}{4}\right),  \quad x<x_1,\\[12pt]
		  \kern-6pt & C_3e^{-\frac{i\pi}{4}}\sqrt{\frac{\mathcal{A}(x)}{|p(x)|}}\exp{\left(-\frac{1}{\hbar}\int^x_{x_2} |p(\xi)|\mathrm{d}\xi\right)}, \quad x_1<x<x_2,\\[12pt]
        \kern-6pt & C_3\sqrt{\frac{\mathcal{A}(x)}{p(x)}}\exp{\left(\frac{i}{\hbar}\int^x_{x_2} p(\xi) \mathrm{d}\xi\right)}, \quad x_2<x.
		\end{cases}
        \label{eq-F123}
\end{equation}
\end{widetext}
Here, the quantity $\gamma$ is defined by
\begin{equation}\label{eq-gamma}
    \gamma=\frac{1}{\hbar}\int^{x_2}_{x_1} |p(\xi)|\mathrm{d}\xi.
\end{equation}
The detailed derivations of the expressions for the upper component in different regions and the associated connection formulas leading to Eq.~\eqref{eq-F123} are given in Appendices~\ref{app-1} and \ref{app-2}. The oscillatory form of $F(x)$ in region I ($x<x_1$) is analogous to the asymptotic behavior of the Airy function in the nonrelativistic case~\cite{Griffiths_Schroeter_2018}.

To derive the transmission probability from the WKB wave functions, we consider the probability current associated with the Dirac equation~\cite{Greiner2003},
\begin{equation}
    j = \Psi^{\dagger}\alpha\Psi
           = F^{*}G+FG^{*}.   \label{eq-jDirac}
\end{equation}
Applying this definition to the transmitted wave in region III from Eq.~\eqref{eq-psi-fg} yields the current
\begin{equation}
    j_3=2|C_3|^2.
\end{equation}
Similarly, for the incident-wave component from region I, which corresponds to the second term in Eq.~\eqref{eq-F1fromx2}, the current becomes
\begin{equation}
    j_1=2|C_3|^2e^{2\gamma}.
\end{equation}
The transmission probability is thus given by
\begin{equation} \label{eq-TDirac}
\begin{split}
    T=\frac{j_3}{j_1}=&\exp{\left(-\frac{2}{\hbar}\int^{x_2}_{x_1} \left|p(\xi)\right|\mathrm{d}\xi\right)} \\
        =&\exp{\left(-\frac{2}{\hbar}\int^{x_2}_{x_1} \sqrt{2m |E_{\text{eff}} - U_{\text{eff}}(\xi)|}  \mathrm{d}\xi\right)}.
\end{split}
\end{equation}
Structurally, Eq. \eqref{eq-TDirac} shares the familiar exponentially suppressed form found in the nonrelativistic WKB approximation~\cite{Landau1958, Griffiths_Schroeter_2018}. However, the relativistic dynamical input is fundamentally different: the local momentum is governed by the effective potential $U_{\text{eff}}(x)$ rather than by a standard Schr\"odinger potential of the form $V(x)$. In the nonrelativistic case, in contrast, one has $p(x)=\sqrt{2m|\epsilon-V(x)|}$ with the kinetic energy $\epsilon=E-m$. Finally, it should be noted that Eq.~\eqref{eq-TDirac} remains valid only when $T\ll1$, consistent with the fact that the WKB approximation inherently requires the reflection probability to be equal to unity.

Physically, a continuously varying potential barrier can be fundamentally discretized into successive rectangular potential barriers. Therefore, deriving the exact transmission probability through a single rectangular barrier provides a natural conceptual bridge to the relativistic WKB transmission probability.
For a rectangular scalar-vector barrier, the wave functions in the three regions can be determined exactly. In this case, the relativistic modifications to the transmission probability arise not only from the definition of the local momentum but also from the spinor matching conditions at the discontinuous boundaries. This exact treatment serves as a useful complement to the smooth-barrier WKB result, and the detailed derivations are given in Appendix~\ref{app-3}. 

The present framework is not restricted to tunneling problems. When applied to an effective potential well, it naturally yields the corresponding bound-state quantization condition, as detailed in Appendix~\ref{app-4}. This broad applicability further establishes $U_{\text{eff}}(x)$ as the fundamental effective potential governing the semiclassical treatment of both the scattering and bound states in the 1+1 dimensional Dirac equation.

\subsection{Half-life of proton radioactivity}
Although this $1+1$ dimensional framework cannot fully describe the proton emission, the effective potential $U_{\text{eff}}(x)$ derived within the framework captures the core relativistic effects. By extending this formalism from the one dimensional coordinate $x$ to the radial coordinate $r$, we are able to practically investigate these relativistic dynamical effects in realistic proton emission processes.

The half-life of proton emission is expressed as
\begin{equation}
T_{1/2} = \frac{\ln 2}{S_p \nu P},   
\end{equation}
where $S_p$, $\nu$, and $P$ denote the spectroscopic factor, the assault frequency, and the penetration probability, respectively.

In previous studies~\cite{Basu200511, Dong200905, 2011-Sahu-Phys.Rev.C, Sahoo201904, Zhu202611}, the penetration probability $P$ is typically evaluated within the WKB approximation as
\begin{equation} \label{eq-Pwkb}
    P = \exp \left( -\frac{2}{\hbar} \int_{r_2}^{r_3} \sqrt{2\mu |Q_p - V_{\text{total}}(r)|}  \mathrm{d}r  \right).  
\end{equation}
Here, $V_{\text{total}}(r)$ denotes the total potential barrier, $r$ is the center-of-mass distance between the emitted proton and the daughter nucleus, $r_2$ and $r_3$ are the second and third classical turning points of $V_{\text{total}}(r)$, $\mu$ is the reduced mass of the proton-daughter nucleus system, and $Q_p$ is the decay energy. The total potential is written as
\begin{equation}
    V_{\text{total}}(r) = V_N(r) + V_C(r) + V_l(r),    
\end{equation}
with $V_N(r)$, $V_C(r)$, and $V_l(r)$ representing the nuclear, Coulomb, and centrifugal contributions, respectively. With the Langer modification \cite{1995-Morehead-J.Math.Phys.}, the centrifugal potential $V_l(r)$ reads
\begin{equation}
    V_l(r)=\frac{(l+\frac{1}{2})^2\hbar^2}{2\mu r^2},
\end{equation}
with $l$ being the orbital angular momentum quantum number of the emitted proton. For a uniformly charged sphere, the Coulomb potential is given by
\begin{equation} \label{eq-Vc}
    V_C(r)=
    \begin{cases}
        \dfrac{Z_de^{2}}{8\pi\varepsilon_{0}R_{c}}
        \left(3-\dfrac{r^{2}}{R_{c}^{2}}\right), & r\le R_{c},\\[8pt]
        \dfrac{Z_de^{2}}{4\pi\varepsilon_{0}r}, & r>R_{c},
    \end{cases}
\end{equation}
where $Z_d$ is the charge number of the daughter nucleus and $R_c$ is the corresponding charge radius. 
Following Ref.~\cite{Sheng2015}, we parameterize the charge radius of the daughter nucleus as
\begin{equation}
R_c = r_c A_d^{1/3},
\end{equation}
where $A_d = A-1$ is the mass number of the daughter nucleus, with $A$ being the mass number of the parent proton emitter, and $r_c = 1.2247$~fm is the charge radius parameter.

For the nuclear part of the proton-emission barrier, we consider two alternative treatments. The first is the conventional choice
\begin{equation}
    V_N(r) = S(r) + V(r),
\end{equation}
while the second is the effective potential derived from our WKB analysis,
\begin{equation} \label{eq-VasUeff}
     V_N(r) = U_{\text{eff}}(r) = S(r) + \frac{E}{m}V(r)+\frac{S^2(r)-V^2(r)}{2m}.
\end{equation}
Since both treatments are constructed from the same scalar and vector potentials, any deviation between them directly reflects the relativistic dynamical effect associated with the choice of the nuclear potential.

Following Ref.~\cite{1991-Koepf-ZPA}, we parameterize the scalar and vector potentials using two Woods-Saxon potentials, $U(r)$ and $W(r)$,
\begin{subequations}
\begin{align}
    S(r)=&\frac{1}{2}[U(r)-W(r)],\\
    V(r)=&\frac{1}{2}[U(r)+W(r)],
\end{align}
\end{subequations}
with
\begin{subequations}
\begin{align}
    U(r) &= \frac{V_{p}}{1 + \exp\left[ (r - R^{p}_{1}) / a^{p}_{1} \right]}, \\
    W(r) &= \frac{-\lambda_{p} V_{p}}{1 + \exp\left[ (r - R_{2}^{p}) / a_{2}^{p} \right]}.
\end{align}
\end{subequations}
The strength parameter $V_p$ smoothly depends on the proton number $Z$ and the neutron number $N$ of the proton emitter,
\begin{equation}
    V_{p} = V^{0}_p \left( 1 + \kappa \frac{N - Z}{N+Z} \right),
\end{equation}
and the radius parameters are taken in the standard form:
\begin{subequations}
\begin{align}
    R_{1}^{p} &= r_{1}^{p} A^{1/3}, \\
    R_{2}^{p} &= r_{2}^{ p} A^{1/3}.
\end{align}
\end{subequations}
The parameters $V_p^{0},\kappa$, $\lambda_p$, $r_1^{p}$, $r_2^{p}$, $a_1^{p}$, and $a^{p}_2$ are assumed to be independent of the nucleus and are shown in Table \ref{tab-1}.

\begin{table}[htbp]
\caption{Nuclear potential parameters from Ref.~\cite{1991-Koepf-ZPA}.}
\label{tab-1}
\renewcommand{\arraystretch}{1.25}
\begin{tabular*}{\columnwidth}{@{\extracolsep{\fill}}ccccccc}
    \hline\hline
    $V^{0}_p$ (MeV) & $\kappa$ & $\lambda_p$ & $r^{p}_1$ (fm) & $a^{p}_1$ (fm) & $r_2^{p}$ (fm) & $a_2^{p}$ (fm) \\
    \hline
    $-71.28$ & 0.462 & 8.97 & 1.250 & 0.612 & 1.140 & 0.647 \\
    \hline\hline
\end{tabular*}
\renewcommand{\arraystretch}{1.0}
\end{table}

In addition to the penetration probability, the half-life also depends on the assault frequency. We therefore need an explicit expression for $\nu$ within the same semiclassical framework. Following Refs.~\cite{Gurvitz1987, yang2023, Lu2024}, we express the assault frequency $\nu$ as
\begin{equation} \label{eq-nu}
    \nu=\mathcal{N}\frac{\hbar}{4\mu},
\end{equation}
where $\mathcal{N}$ is the quasi-classical normalization factor. In Ref.~\cite{Gurvitz1987}, the analytical expression for this factor was simplified by applying the bound-state quantization condition to alter the integration limits. However, since the decay energy $Q_p$ in proton emission does not correspond to a true bound-state eigenenergy, this quantization condition is strictly inapplicable. We therefore adopt the original normalization condition

\begin{equation} \label{eq-Nmod}
    \mathcal{N}\int_{r_1}^{r_2}\frac{1}{k(r)} \cos^2\left( \int_{r}^{r_2} k(\xi)  \dd \xi  
    - \frac{\pi}{4}\right) \dd r = 1,
\end{equation}
where 
\begin{equation}\label{eq-kr}
    k(r) = \frac{1}{\hbar}\sqrt{2\mu[Q_p - V_{\text{total}}(r)]},
\end{equation}
with $r_1$ and $r_2$ denoting the inner two classical turning points of $V_{\text{total}}(r)$. A detailed derivation of Eq.~\eqref{eq-Nmod} is provided in Appendix~\ref{app-2}.

\section{Numerical details}\label{sec4}

Since our purpose here is not to reproduce the experimental half-lives as accurately as possible, but rather to isolate the relativistic dynamical effect associated with the choice of the nuclear potential, we set the spectroscopic factor to $S_p=1$ throughout this work. In evaluating $U_{\text{eff}}(r)$ via Eq.~\eqref{eq-VasUeff}, the single-particle energy is taken as $E= Q_p+m$, with the proton mass $m=938.5$\ MeV. The reduced mass $\mu$ is conventionally approximated as $\mu\approx\frac{A-1}{A}\cdot m$. 

In the calculation of the assault frequency, the normalization integral in Eq.~\eqref{eq-Nmod} exhibits integrable endpoint singularities at the turning points $r_1$ and $r_2$. For example, near $r =r_1$ one has
\begin{equation}
     \frac{\dd r}{k(r)}\propto\frac{\dd r}{\sqrt{r-r_1}},\qquad r\rightarrow r_1^{+}.
\end{equation}
To regularize the numerical integration, we introduce the transformation
\begin{equation}
    r = r_1 + (r_2-r_1) \sin^2\theta, \qquad  \theta \in[0,\pi/2].
\end{equation} 
Under this substitution, in the neighborhood of $r_1$, the integration variable changes from $r$ to $\theta$, yielding
\begin{equation}
    \frac{\dd r}{\sqrt{r-r_1}}
    =2\sqrt{r_2-r_1}\cos\theta\,\dd\theta,
    \qquad \theta\to 0^{+},
\end{equation}
which ensures that the transformed integrand remains finite at the endpoint.

\section{Results and Discussion}\label{sec5}
After systematically calculating the proton emission properties for a wide range of nuclei, we deliberately selected $^{144}\text{Tm}$, $^{155}\text{Ta}$, and $^{170}\text{Au}$ as representative examples because they cover different magnitudes of relativistic dynamical corrections to the half-life, from highly pronounced to relatively mild.

\begin{figure}[!ht]
    \centering
	\includegraphics[width=0.95\linewidth]{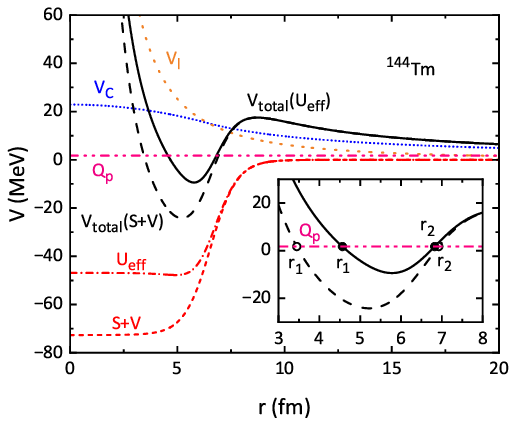}
	\caption{Proton-emission potential barrier for $^{144}\mathrm{Tm}$. The nuclear component is taken as either $U_{\text{eff}}$ (red dash-dotted line) or $S+V$ (red dashed line), together with the Coulomb component $V_C$ (blue short-dotted line) and the Langer-modified centrifugal term $V_l$ (orange dotted line) for $l=5$. The resulting total barriers $V_{\text{total}}(U_{\text{eff}})$ and $V_{\text{total}}(S+V)$ are shown by the black solid and black dashed lines, respectively. The experimental decay energy $Q_p=1.712$ MeV is indicated by the horizontal pink line. Inset: turning points $r_1$ and $r_2$ for the two choices of the nuclear potential.}
    \label{fig2}
\end{figure}

Using the two nuclear potentials, $U_{\text{eff}}(r)$ in Eq.~\eqref{eq-VasUeff} and the conventional choice $S(r)+V(r)$, we first examine the resulting total potential barriers for $^{144}\mathrm{Tm}$ as an example, as shown in Fig.~\ref{fig2}. In the following, $X(U_{\text{eff}})$ and $X(S+V)$ denote the results obtained with $U_{\text{eff}}$ and $S+V$, respectively, for any quantity $X$. Figure \ref{fig2} shows that $U_{\text{eff}}$ exceeds $S+V$ by about 25.8 MeV throughout the nuclear interior, thereby producing a shallower potential well. This difference modifies the turning-point structure of the total barrier: the first turning point $r_1$ moves outward, whereas the second turning point $r_2$ moves inward when $U_{\text{eff}}(r)$ is used instead of $S(r)+V(r)$, as indicated in the inset. However, in the nuclear exterior, the two total barriers become nearly identical, and the third turning point $r_3$ is therefore unchanged.

\begin{figure}[!ht]
    \centering
	\includegraphics[width=0.95\linewidth]{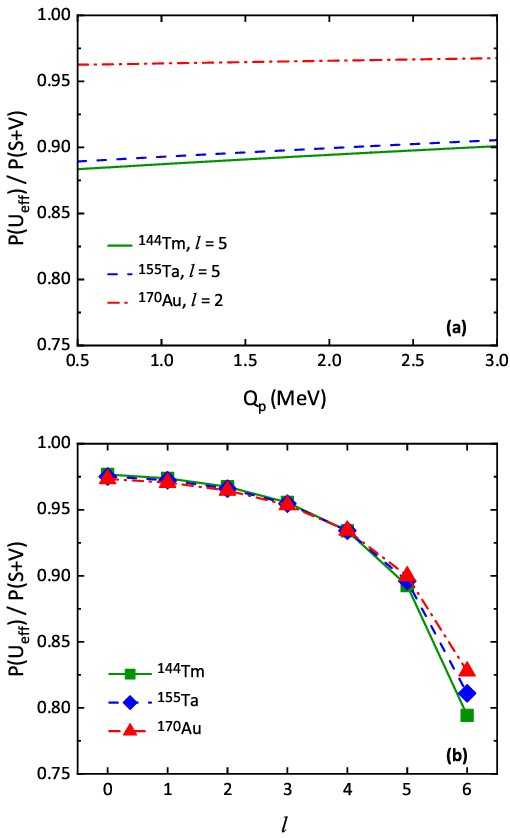}
    \caption{Dependence of $P(U_{\mathrm{eff}})/P(S+V)$ on the decay energy $Q_p$ and the orbital angular momentum $l$. Panel (a) shows the variation with $Q_p$ at fixed $l$ values, and panel (b) shows the variation with $l$ at fixed experimental decay energies. The results are shown for $^{144}\text{Tm}$ (green solid line or square), $^{155}\text{Ta}$ (blue dashed line or diamond), and $^{170}\text{Au}$ (red dash-dotted line or triangle). The experimental $Q_p$ values are accessed via \cite{AME2020} and the deduced $l$ values are taken from Ref.~\cite{zhang2024}.}
    \label{fig3}
\end{figure}

Figure~\ref{fig3} shows that replacing $S(r)+V(r)$ with $U_{\text{eff}}(r)$ leads to a systematic reduction of the penetration probability for all three proton emitters considered here. In Fig.~\ref{fig3}(a), the quantity $P(U_{\mathrm{eff}})/P(S+V)$, which directly characterizes the relativistic dynamical effect on the penetration probability, remains less than unity throughout the whole $Q_p$ range. This implies that $P(U_{\mathrm{eff}})<P(S+V)$ in all cases. Such a relativistic dynamical effect arises because $U_{\text{eff}}$ modifies the inner part of the barrier and shifts the second turning point inward, i.e., $r_2(U_{\text{eff}})<r_2(S+V)$, thereby increasing the WKB action in Eq.~\eqref{eq-Pwkb} and reducing the tunneling probability. Although the magnitude of the effect depends on the nucleus, its dependence on $Q_p$ is rather weak, as indicated by the nearly flat curves. Quantitatively, for $^{170}$Au, the ratio remains above 0.95, indicating a more modest suppression of less than $5\%$. In contrast, the ratio $P(U_{\mathrm{eff}})/P(S+V)$ drops to roughly 0.89--0.90 for $^{144}$Tm, which directly corresponds to a $10\%$--$11\%$ reduction in the penetration probability.

More importantly, Fig.~\ref{fig3}(b) reveals a clear dependence of the relativistic dynamical effect on the orbital angular momentum. For all three nuclei, the ratio $P(U_{\mathrm{eff}})/P(S+V)$ is less than unity and decreases monotonically with $l$ over the range considered, reaching its minimum at $l=6$. In particular, the value drops to about $79.4\%$ for $^{144}$Tm at $l=6$. This trend indicates that high-$l$ proton emission is especially sensitive to the choice of the nuclear potential. 

\begin{figure}[!ht]
    \centering
	\includegraphics[width=0.95\linewidth]{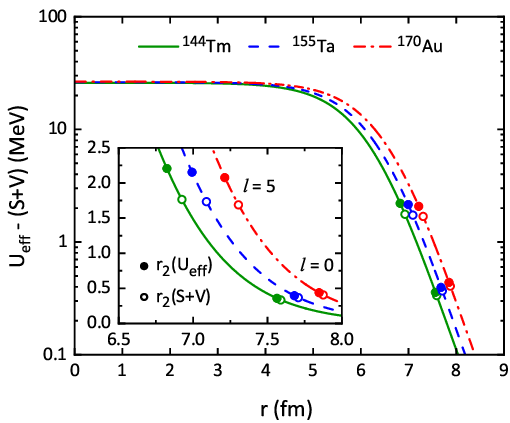}
    \caption{Difference $U_{\text{eff}}-(S+V)$ in the nuclear interior for $^{144}$Tm (green solid line), $^{155}$Ta (blue dashed line), and $^{170}$Au (red dash-dotted line), evaluated at their experimental $Q_p$ values. Inset: the second turning points $r_2$ for $U_{\text{eff}}$ (solid circles) and $S+V$ (open circles) with $l=0$ and $5$.}
    \label{fig4}
\end{figure}

The trend in Fig.~\ref{fig3}(b) can be traced back to the behavior of the second turning point $r_2$ in Fig.~\ref{fig4}. Because $U_{\text{eff}}-(S+V)$ is positive throughout the nuclear interior, one has $r_2(U_{\text{eff}})<r_2(S+V)$. The inset of Fig.~\ref{fig4} further shows that this separation increases with $l$. Since the outer turning point $r_3$ is almost unchanged, the inward shift of $r_2(U_{\text{eff}})$ increases the WKB action and hence suppresses $P(U_{\text{eff}})$ more strongly at higher orbital angular momentum. In this way, the stronger relativistic dynamical effect at large $l$ follows directly from the turning-point structure of the barrier.

\begin{figure}[!ht]
    \centering
	\includegraphics[width=0.95\linewidth]{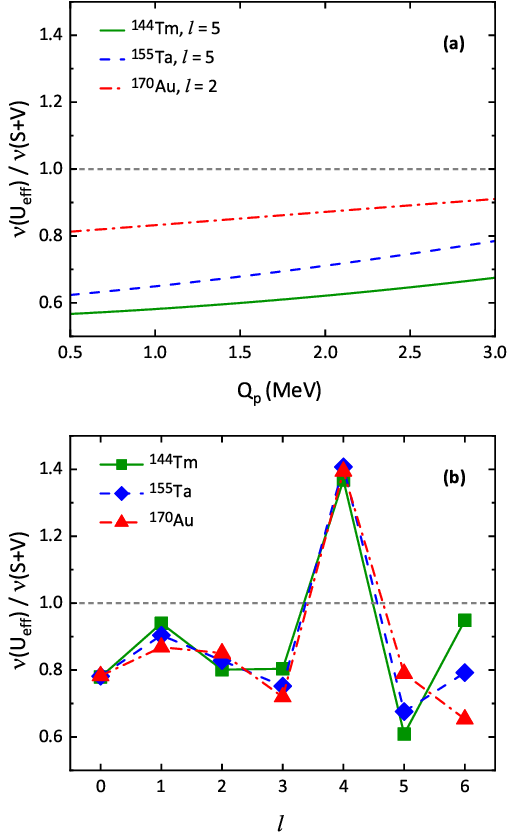}
    \caption{Same as Fig.~\ref{fig3}, but for the quantity $\nu(U_{\mathrm{eff}})/\nu(S+V)$.}
    \label{fig5}
\end{figure}

Figure~\ref{fig5} displays the relativistic dynamical effect on the assault frequency through the quantity $\nu(U_{\mathrm{eff}})/\nu(S+V)$. In Fig.~\ref{fig5}(a), this quantity remains less than unity over the whole $Q_p$ range for all three nuclei, indicating that the assault frequency is generally reduced when $U_{\text{eff}}$ is used. Among the three nuclei shown, $^{144}$Tm exhibits the most pronounced relativistic suppression; its frequency ratio drops to approximately 0.57--0.68, which yields a substantial reduction of $32\%$--$43\%$. In addition, all three curves increase with increasing $Q_p$, which suggests that the relativistic dynamical effect on the assault frequency becomes weaker at higher decay energies.

In contrast to the monotonic behavior found for the penetration probability in Fig.~\ref{fig3}(b), Fig.~\ref{fig5}(b) exhibits a pronounced oscillatory dependence on the orbital angular momentum. For most $l$ values, $\nu(U_{\mathrm{eff}})/\nu(S+V)$ remains less than unity. At $l=4$, however, it becomes greater than unity for all three nuclei, implying an enhancement of the assault frequency when $U_{\text{eff}}$ is adopted. In particular, the value for $^{144}\text{Tm}$ reaches about $1.4$, showing that the relativistic dynamical effect on the assault frequency can be both strong and nonmonotonic.

\begin{figure}[t]
    \centering
    \includegraphics[width=0.95\linewidth]{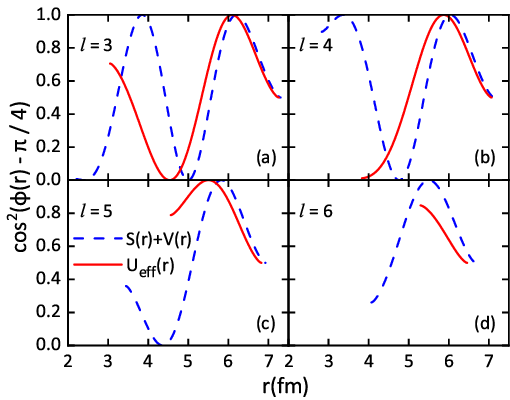}
    \caption{Behavior of $\cos^2[\phi(r)-\pi/4]$ in Eq.~\eqref{eq-Nmod} between the turning points $r_1$ and $r_2$ for $^{144}\text{Tm}$ at different $l$ values, where $\phi(r)=\int_r^{r_2} k(\xi)\dd\xi$. Panels (a)–(d) correspond to $l=3$, 4, 5, and 6, respectively. The red solid and blue dashed lines represent the results obtained with $U_{\text{eff}}(r)$ and $S(r)+V(r)$, respectively. The experimental $Q_p$ value is employed for all cases.}
    \label{fig6}
\end{figure}

This oscillatory behavior can be traced back to the cosine factor in the normalization condition~\eqref{eq-Nmod}. As shown in Fig.~\ref{fig6} for $^{144}\mathrm{Tm}$, this factor is not a rapidly oscillating function between $r_1$ and $r_2$. It therefore cannot be replaced by its average value $1/2$, and the assault frequency remains sensitive to the detailed phase structure of the wave function in the inner region. The pronounced difference between the $l=4$ and $l=5$ results thus follows directly from the exact treatment of this factor, which explains the nonmonotonic relativistic dynamical effect observed in Fig.~\ref{fig5}(b).

\begin{figure}[!ht]
    \centering
    \includegraphics[width=0.95\linewidth]{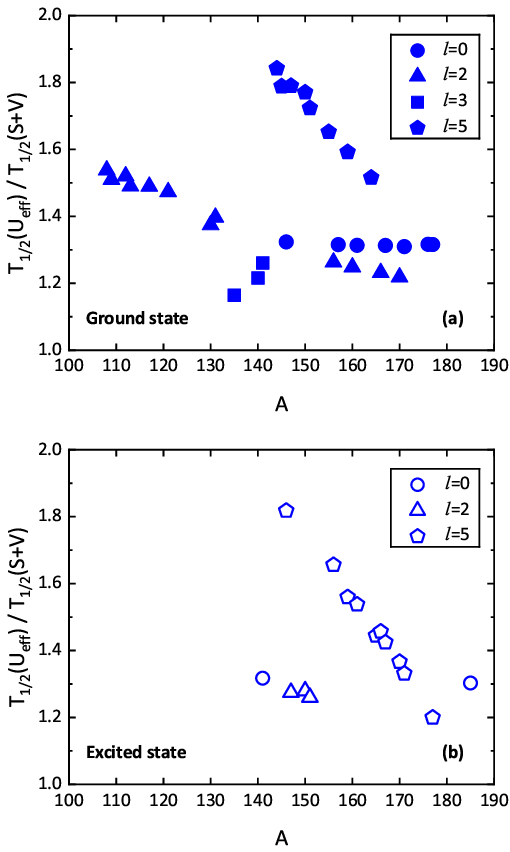}
    \caption{Half-life ratio $T_{1/2}(U_{\text{eff}})/T_{1/2}(S+V)$ as a function of mass number for ground-state proton emitters in panel (a) and excited-state proton emitters in panel (b). Filled symbols denote ground states with $l=0$, 2, 3, and 5 (circles, triangles, squares, and pentagons, respectively), while open symbols denote excited states with $l=0$, 2, and 5. The mass numbers and deduced orbital angular momenta are taken from Ref.~\cite{zhang2024}.}
    \label{fig7}
\end{figure}

After combining the relativistic corrections to both the penetration probability and the assault frequency, we finally examine their net impact on the half-life. Figure~\ref{fig7} displays the ratio $T_{1/2}(U_{\text{eff}})/T_{1/2}(S+V)$ as a function of mass number and orbital angular momentum for ground-state and excited-state proton emitters. In all cases, the ratio remains above unity, showing that the use of $U_{\text{eff}}$ systematically leads to longer half-lives.

For the ground-state emitters shown in Fig.~\ref{fig7}(a), the ratios for nuclei with $l=0$ and 3 are typically in the range $1.2$--$1.4$, corresponding to half-life increases of $20\%$--$40\%$. A stronger mass-number dependence appears for $l=2$ and 5, with the ratio increasing toward lighter nuclei. The largest values reach about $1.54$ for $A=108$ with $l=2$ and $1.84$ for $A=144$ with $l=5$, corresponding to relativistic enhancements of $54\%$ and $84\%$, respectively.

\begin{table*}[htbp]
\caption{Ratios characterizing the relativistic dynamical effects on the penetration probability, assault frequency, and half-life of proton emitters. The experimental decay energies for the ground states are adopted from the Atomic Mass Evaluation (AME2020)~\cite{AME2020}, where values appended with '\#' (specifically for $^{130}$Eu, $^{159}$Re, and $^{164}$Ir) indicate estimated values rather than direct experimental measurements. For the isomeric states, the experimental decay energies are taken from Ref.~\cite{Xiao_2023}. The deduced $l$ values are obatined from Ref.~\cite{zhang2024}.}
\label{tab-2}
\renewcommand{\arraystretch}{1.3} %
\begin{tabular*}{\textwidth}{@{\extracolsep{\fill}}cccccccccccc}
\hline\hline
\noalign{\vspace{4pt}} %
Nucleus & $Q_p$ (MeV) & $l$ & $\dfrac{P(U_\text{eff})}{P(S+V)}$ & $\dfrac{\nu(U_\text{eff})}{\nu(S+V)}$ & $\dfrac{T_{1/2}(U_\text{eff})}{T_{1/2}(S+V)}$ & Nucleus & $Q_p$ (MeV) & $l$ & $\dfrac{P(U_\text{eff})}{P(S+V)}$ & $\dfrac{\nu(U_\text{eff})}{\nu(S+V)}$ & $\dfrac{T_{1/2}(U_\text{eff})}{T_{1/2}(S+V)}$ \\
\noalign{\vspace{4pt}} %
\hline
$^{108}$I & 0.597 & 2 & 0.9703 & 0.6701 & 1.5381 & $^{156}$Ta & 1.020 & 2 & 0.9655 & 0.8200 & 1.2630 \\
$^{109}$I & 0.820 & 2 & 0.9710 & 0.6821 & 1.5098 & $^{156m}$Ta & 1.110 & 5 & 0.8950 & 0.6748 & 1.6558 \\
$^{112}$Cs & 0.816 & 2 & 0.9699 & 0.6782 & 1.5203 & $^{157}$Ta & 0.935 & 0 & 0.9747 & 0.7798 & 1.3156 \\
$^{113}$Cs & 0.973 & 2 & 0.9705 & 0.6915 & 1.4900 & $^{159}$Re & 1.600\# & 5 & 0.8967 & 0.7005 & 1.5921 \\
$^{117}$La & 0.820 & 2 & 0.9695 & 0.6926 & 1.4893 & $^{159m}$Re & 1.801 & 5 & 0.8980 & 0.7140 & 1.5596 \\
$^{121}$Pr & 0.890 & 2 & 0.9688 & 0.7007 & 1.4730 & $^{160}$Re & 1.267 & 2 & 0.9652 & 0.8300 & 1.2482 \\
$^{130}$Eu & 1.530\# & 2 & 0.9688 & 0.7511 & 1.3742 & $^{161}$Re & 1.197 & 0 & 0.9743 & 0.7815 & 1.3134 \\
$^{131}$Eu & 0.947 & 2 & 0.9680 & 0.7395 & 1.3969 & $^{161m}$Re & 1.317 & 5 & 0.8977 & 0.7244 & 1.5377 \\
$^{135}$Tb & 1.188 & 3 & 0.9551 & 0.8994 & 1.1642 & $^{164}$Ir & 1.560\# & 5 & 0.8978 & 0.7349 & 1.5156 \\
$^{140}$Ho & 1.094 & 3 & 0.9546 & 0.8618 & 1.2156 & $^{165m}$Ir & 1.711 & 5 & 0.9001 & 0.7691 & 1.4445 \\
$^{141}$Ho & 1.177 & 3 & 0.9553 & 0.8304 & 1.2606 & $^{166}$Ir & 1.152 & 2 & 0.9647 & 0.8417 & 1.2316 \\
$^{141m}$Ho & 1.247 & 0 & 0.9769 & 0.7770 & 1.3174 & $^{166m}$Ir & 1.332 & 5 & 0.8990 & 0.7635 & 1.4568 \\
$^{144}$Tm & 1.712 & 5 & 0.8923 & 0.6085 & 1.8418 & $^{167}$Ir & 1.070 & 0 & 0.9737 & 0.7822 & 1.3130 \\
$^{145}$Tm & 1.736 & 5 & 0.8942 & 0.6253 & 1.7885 & $^{167m}$Ir & 1.245 & 5 & 0.8998 & 0.7802 & 1.4245 \\
$^{146}$Tm & 0.896 & 0 & 0.9758 & 0.7743 & 1.3235 & $^{170}$Au & 1.472 & 2 & 0.9645 & 0.8509 & 1.2184 \\
$^{146m}$Tm & 1.206 & 5 & 0.8922 & 0.6166 & 1.8178 & $^{170m}$Au & 1.752 & 5 & 0.9014 & 0.8120 & 1.3662 \\
$^{147}$Tm & 1.059 & 5 & 0.8929 & 0.6257 & 1.7899 & $^{171}$Au & 1.448 & 0 & 0.9734 & 0.7845 & 1.3095 \\
$^{147m}$Tm & 1.120 & 2 & 0.9671 & 0.8113 & 1.2745 & $^{171m}$Au & 1.702 & 5 & 0.9023 & 0.8324 & 1.3314 \\
$^{150}$Lu & 1.270 & 5 & 0.8929 & 0.6327 & 1.7703 & $^{176}$Tl & 1.265 & 0 & 0.9725 & 0.7811 & 1.3165 \\
$^{150m}$Lu & 1.290 & 2 & 0.9663 & 0.8089 & 1.2794 & $^{177}$Tl & 1.156 & 0 & 0.9725 & 0.7814 & 1.3159 \\
$^{151}$Lu & 1.241 & 5 & 0.8943 & 0.6488 & 1.7234 & $^{177m}$Tl & 1.963 & 5 & 0.9059 & 0.9200 & 1.1998 \\
$^{151m}$Lu & 1.301 & 2 & 0.9666 & 0.8211 & 1.2599 & $^{185m}$Bi & 1.607 & 0 & 0.9731 & 0.7887 & 1.3029 \\
$^{155}$Ta & 1.453 & 5 & 0.8958 & 0.6758 & 1.6519 & & & & & & \\
\hline\hline
\end{tabular*}
\end{table*}

In Fig.~\ref{fig7}(b), the excited-state emitters exhibit a similar pattern. The ratios again remain above unity throughout, while the $l=5$ nuclei show the strongest dependence on mass number, reaching a maximum value of about $1.82$ at $A=146$. In contrast, the nuclei with $l=0$ and 2 remain in the narrower range $1.2$--$1.4$. Overall, these results show that the relativistic dynamical effect on the half-life is most pronounced for lighter nuclei and for proton emission with larger orbital angular momentum, especially $l=5$.

Table~\ref{tab-2} summarizes the ratios of the penetration probability, assault frequency, and half-life for all proton emitters included in Fig.~\ref{fig7}. In every case, both $P(U_{\text{eff}})/P(S+V)$ and $\nu(U_{\text{eff}})/\nu(S+V)$ are smaller than unity, which means that $U_{\text{eff}}$ systematically suppresses the penetration probability and the assault frequency, and hence increases the half-life. It is observed that, with the notable exception of $^{177m}$Tl, the ratio $\nu(U_{\text{eff}})/\nu(S+V)$ is systematically smaller than $P(U_{\text{eff}})/P(S+V)$. This indicates that the relativistic dynamical effect is generally stronger for the assault frequency than for the penetration probability.

\section{SUMMARY}\label{sec6}

Within a relativistic framework, the effective nuclear potential and the penetration probability for proton emission are derived from the 1+1 dimensional Dirac equation via the WKB approximation. The resulting penetration probability retains the familiar exponential form and is formally analogous to that obtained from the radial Schr\"odinger equation, while the semiclassical momentum is determined by the Schr\"odinger-equivalent potential $U_{\text{eff}}(r)$ rather than by the simple sum of scalar and vector potentials $S(r)+V(r)$. The same framework also reproduces the Bohr--Sommerfeld quantization condition and the transmission probability for a rectangular potential barrier.

The relativistic dynamical effects are quantified through a comparison between calculations based on $U_{\text{eff}}(r)$ and $S(r)+V(r)$. The penetration probability obtained with $U_{\text{eff}}(r)$ is systematically reduced. This reduction shows only a weak dependence on the decay energy, but becomes more pronounced with increasing orbital angular momentum $l$, indicating that relativistic dynamical effects are particularly important for high-$l$ proton emission.

The assault frequency also exhibits noticeable relativistic dynamical effects. In most cases, the use of $U_{\text{eff}}(r)$ leads to a reduction in the assault frequency. However, for several nuclei, including $^{144}\mathrm{Tm}$, $^{155}\mathrm{Ta}$, and $^{170}\mathrm{Au}$ with $l=4$, an enhancement is obtained instead. This behavior originates from the cosine factor entering the normalization condition~\eqref{eq-Nmod}.

As a consequence of the combined changes in the penetration probability and assault frequency, the half-lives calculated with $U_{\text{eff}}(r)$ are longer for all proton emitters considered in this work. In particular, the half-life of $^{144}\mathrm{Tm}$ increases by as much as $84\%$. Overall, with the rare exception of $^{177m}$Tl, the relativistic correction to the assault frequency is found to be more significant than that to the penetration probability for the nuclei studied here. The present work provides a benchmark for improving the self-consistency of descriptions of proton radioactivity in relativistic WKB frameworks, and serves as a useful starting point for extending the analysis to the 3+1 dimensional Dirac equation.

\begin{acknowledgments}
This research was supported by the Super Computing Center of Beijing Normal University. 
S.W. acknowledges funding from the China Scholarship Council (CSC) (File No.~[202406050068]), the National Natural Science Foundation of China under grant No.~12575130, and the Chongqing Natural Science Foundation under grant No.~CSTB2025NSCQ-GPX0742. 
H.L. acknowledges funding from the JSPS Grant-in-Aid under Grants Nos. 26K07063 and 26K01431. 

\end{acknowledgments}

\appendix

\begin{widetext}

\section{Connection formulas between regions III and II}\label{app-1}

To derive the connection formula across the turning point $x_2$ in Fig.~\ref{fig1}, we extend the real coordinate $x$ to the complex variable $z$. In the proton-emission problem, only an outgoing wave exists in region III ($x > x_2$), so the upper component of the spinor $\Psi(x)$ takes the WKB form 
\begin{equation}\label{eq-F3}
    F_3(x) = C_3 \sqrt{\frac{\mathcal{A}(x)}{p(x)}} 
             \exp\!\left( \frac{i}{\hbar} \int_{x_2}^{x} p(\xi) \, \mathrm{d}\xi \right),
\end{equation}
where $\mathcal{A}(x)$ is defined in Eq.~\eqref {eq-Adef} and the lower integration limit is chosen as the turning point $x_2$. To continue this solution into the classically forbidden region II ($x_1 < x < x_2$), where $p(x)$ becomes purely imaginary, we follow the method in Refs.~\cite{zwaan1929,Landau1958,Heading1962} and bypass the turning point $x_2$ in the complex plane.

Near $x = x_2$, the effective potential can be linearized as
\begin{equation} \label{eq-linx2}
    E_{\mathrm{eff}} - U_{\mathrm{eff}}(x) 
    \approx \mathcal{S}_2 (x - x_2), \qquad 
    \mathcal{S}_2 \equiv -\left. \frac{\mathrm{d} U_{\mathrm{eff}}}{\mathrm{d}x} \right|_{x = x_2} > 0.
\end{equation}
Consequently, the local momentum $p(x)$ in region III reduces to
\begin{equation}
    p(x) \approx \sqrt{ 2m \mathcal{S}_2 (x - x_2) }.
\end{equation}  
Additionally, in our present case, $\mathcal{A}(x)$ is strictly positive along the entire real axis.  Since the WKB approximation is valid for a slowly varying and sufficiently wide potential barrier, there exists an overlap region in which both the linear expansion of the effective potential and the WKB approximation remain valid~\cite{Landau1958}. 

\begin{figure}[!ht]
    \centering
    \includegraphics[width=0.40\linewidth]{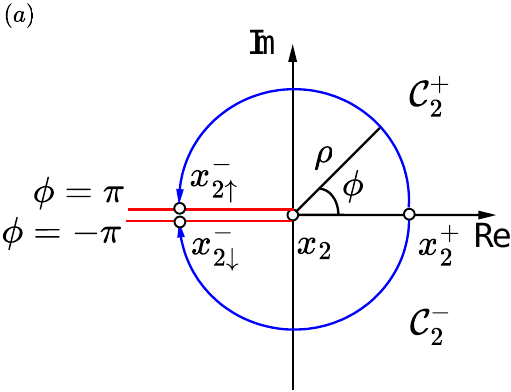}
    \hspace{0.05\linewidth} 
    \includegraphics[width=0.40\linewidth]{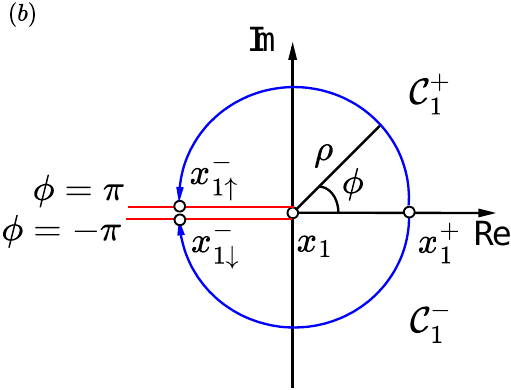}
    
    \caption{Analytic continuation in the complex plane. (a) In the neighborhood of $x_2$. The blue contours $\mathcal{C}^{+}_2$ and $\mathcal{C}^{-}_2$ are semicircles of radius $\rho$ in the complex plane from $x^+_{2}$ to the left of $x_2$. To ensure that $(z-x_2)^{1/2}$ and $(z-x_2)^{-1/4}$ remain single-valued, a branch cut is introduced along the real axis from $x_2$ to $-\infty$. The red lines above and below the axis denote the upper and lower edges ($\phi = \pi,-\pi$) of this cut, respectively. (b) Same as (a), but for the analytic continuation in the neighborhood of $x_1$.}
    \label{fig8}
\end{figure}

We now analytically continue $F_3(x)$ along the upper contour $\mathcal{C}_2^{+}$ depicted in Fig.~\ref{fig8}(a), namely a semicircle of radius $\rho$ in the upper half-plane connecting the two sides of the turning point $x_2$. On this contour, the upper component becomes 
\begin{equation} \label{eq-F3inx2int}
\begin{split}
    F_3(z)=&\ C_3\frac{\left[\mathcal{A}(z)\right]^{\frac{1}{2}}}{\left[2m\mathcal{S}_2(z-x_2)\right]^{\frac{1}{4}}}
        \times\exp{\left(\frac{i}{\hbar}\int^z_{x_2} \sqrt{2m\mathcal{S}_2(\xi -x_2)}\mathrm{d}\xi \right)}.
\end{split}
\end{equation}
Since $z=x_2$ is a branch point, a branch cut in the $z$-plane must be introduced to ensure that $(z - x_2)^{1/2}$ and $(z-x_2)^{-1/4}$ remain single-valued in Eq.~\eqref{eq-F3inx2int}. We choose the cut to extend along the real axis from $x_2$ to $-\infty$, as shown in Fig.~\ref{fig8}(a). The upper and lower edges of the cut correspond to $\phi=\pi$ and $\phi = -\pi$, respectively. Performing the integration in Eq.~\eqref{eq-F3inx2int} yields 
\begin{equation} \label{eq-F3inx2}
\begin{split}
    F_3(z)=\ C_3\frac{\left[\mathcal{A}(z)\right]^{\frac{1}{2}}}{\left[2m\mathcal{S}_2(z-x_2)\right]^{\frac{1}{4}}}\times\exp{\left(\frac{i2\sqrt{2m\mathcal{S}_2}}{3\hbar}(z-x_2)^{\frac{3}{2}}\right)}.
\end{split}
\end{equation}
Along the upper contour $\mathcal{C}^{+}_2$, we parameterize the coordinate as
\begin{equation}
    z-x_2=\rho e^{i\phi}, \qquad \phi \in [0,\pi],
\end{equation}
so that 
\begin{equation}
    i(z-x_2)^{\frac{3}{2}} = i\rho^{3/2}e^{i3\phi/2} 
    = \rho^{3/2}\left(-\sin\frac{3}{2}\phi+i\cos\frac{3}{2}\phi\right).
\end{equation}
At the upper edge of the cut to the left of $x_2$, where $\phi=\pi$, the exponential factor in Eq.~\eqref{eq-F3inx2} becomes purely real:
\begin{equation} \label{eq-pint}
    \frac{2\sqrt{2m\mathcal{S}_2}}{3\hbar}\rho^{3/2}
        =-\frac{1}{\hbar}\int^{x^{-}_{2\uparrow}}_{x_2} \sqrt{2m\mathcal{S}_2(x_2-x)}\mathrm{d}x
        =-\frac{1}{\hbar}\int^{x^{-}_{2\uparrow}}_{x_2} |p(x)|\mathrm{d}x.
\end{equation}
Meanwhile, the prefactor in Eq.~\eqref{eq-F3inx2} becomes
\begin{equation}
    (z-x_2)^{-1/4}\Big|_{z=x^{-}_{2\uparrow}}
    = \left(\rho e^{i\phi}\right)^{-1/4} \Big|_{\phi=\pi}=(x_{2}-x^{-}_{2\uparrow})^{-1/4}e^{-i\pi/4}.
\end{equation}
Therefore, in the classically forbidden region II, the upper component takes the WKB form
\begin{equation} \label{eq-F2fromx3up}
    F_2^{+}(x)=C_2 \sqrt{\frac{\mathcal{A}(x)}{|p(x)|}}
        \exp{\left(+\frac{1}{\hbar}\int^{x_2}_{x} |p(\xi) |\mathrm{d}\xi \right)},
    \qquad x_1<x<x_2,
\end{equation}
where the integration constant $C_2$ is related to $C_3$ by 
\begin{equation} \label{eq-C2C3}
    C_2=C_3e^{-i\pi/4}.
\end{equation}
Notably, this specific branch of the solution represents an exponentially growing wave when moving inward from the outer turning point $x_2$ toward the inner turning point $x_1$.

The analytic continuation can also be carried out along the lower contour $\mathcal{C}_2^{-}$ depicted in Fig.~\ref{fig8}(a), namely a semicircle of radius $\rho$ in the lower half-plane connecting the two sides of the turning point $x_2$. In this case, at the lower edge of the cut where $\phi\to-\pi$, the continuation generates another solution in region II,
\begin{equation}
    F_2^{-}(x) = C'_2 \sqrt{\frac{\mathcal{A}(x)}{|p(x)|}} \exp\!\left(-\frac{1}{\hbar}\int_x^{x_2}|p(\xi)|\,\mathrm{d}\xi\right),
    \qquad x_1 < x < x_2,
\end{equation}
where $C'_2 = C_3 e^{i\pi/4}$. In contrast to the previous branch, this specific solution represents an exponentially decaying wave as $x$ moves inward from the outer turning point $x_2$ toward the inner turning point $x_1$. This contribution is subdominant within the WKB matching procedure. However, the accuracy of the quasi-classical approximation is not sufficient to retain exponentially small terms superimposed on exponentially large ones~\cite{Landau1958, schiff1968quantum}. 
Therefore, only the continuation~\eqref{eq-F2fromx3up} along $\mathcal{C}_2^{+}$ is retained, and Eq.~\eqref{eq-C2C3} gives the relevant connection formula between regions III and II.

\section{Connection formulas between regions II and I}\label{app-2}

Similarly, to obtain the form of the upper component in region I, we apply the same method to continue the WKB solution from region II across the turning point $x_1$. Near $x=x_1$, the effective potential can be linearized as  
\begin{equation} \label{eq-linx1}
    E_{\text{eff}}-U_{\text{eff}}\approx \mathcal{S}_1(x-x_1),\qquad 
    \mathcal{S}_1\equiv -\left.\frac{\dd U_{\text{eff}}}{\dd x}\right|_{x=x_{1}}<0,
\end{equation}
so that in the classically forbidden region II, 
\begin{equation}
    { p(x)\approx \sqrt{2m\mathcal{S}_1(x-x_1)},}
\end{equation}
where $p(x)$ is purely imaginary here.

First, we analytically continue $F_2(x)$ along the upper contour $\mathcal{C}^{+}_1$ depicted in Fig.~\ref{fig8}(b), namely a semicircle of radius $\rho$ in the upper half-plane connecting the two sides of the turning point $x_1$. On this contour, the upper component~\eqref{eq-F2fromx3up} becomes 
\begin{equation}  \label{eq-F2inx1int}
\begin{split}
    F_2(z)&=C_2e^{\gamma}\frac{[\mathcal{A}(z)]^{\frac{1}{2}}}{[2m|\mathcal{S}_1|(z-x_1)]^{\frac{1}{4}}}\times
    \exp{\left(-\frac{1}{\hbar}\int^z_{x_1} \sqrt{2m|\mathcal{S}_1|(\xi -x_1)}\mathrm{d}\xi \right)} ,
\end{split}
\end{equation}
where we have used Eq.~\eqref{eq-gamma} to shift the lower integration limit from $x_2$ to $x_1$. Since $z=x_1$ is again a branch point, we introduce a branch cut in the $z$-plane along the real axis to the left of $x_1$, as shown in Fig.~\ref{fig8}(b). Performing the integration in Eq.~\eqref{eq-F2inx1int} yields
\begin{equation}  \label{eq-F2inx1}
\begin{split}
    F_2(z)
    &=C_2e^{\gamma}\frac{[\mathcal{A}(z)]^{\frac{1}{2}}}{[2m|\mathcal{S}_1|(z-x_1)]^{\frac{1}{4}}}\times
    \exp{\left(-\frac{2\sqrt{2m|\mathcal{S}_1|}}{3\hbar}(z-x_1)^{\frac{3}{2}}\right)}.
\end{split}
\end{equation}
Along the upper contour $\mathcal{C}^{+}_1$, we parameterize the coordinate as
\begin{equation}
    z-x_1=\rho e^{i\phi},\qquad  \phi \in [0, \pi],
\end{equation}
so that 
\begin{equation}
    -(z-x_1)^{\frac{3}{2}} =-\rho ^{3/2}e^{i3\phi/2}= -\rho^{3/2}\left(\cos\frac{3}{2}\phi+i\sin\frac{3}{2}\phi\right).
\end{equation}
At the upper edge of the cut to the left of $x_1$, where $\phi=\pi$, the exponential factor becomes purely imaginary:
\begin{equation}
        \frac{2\sqrt{2m\mathcal{S}_1}}{3\hbar}\rho^{3/2}i
        =-\frac{i}{\hbar}\int^{x^{-}_{1\uparrow}}_{x_1} \sqrt{2m|\mathcal{S}_1|(x_1-\xi )}\mathrm{d}\xi
        =-\frac{i}{\hbar}\int^{x^{-}_{1\uparrow}}_{x_1} p(\xi)\mathrm{d}\xi.
\end{equation}
Meanwhile, the prefactor in Eq.~\eqref{eq-F2inx1} reduces to
\begin{equation}
    (z-x_1)^{-1/4}\Big|_{z=x_{1\uparrow}^{-}}
    =(\rho e^{i\phi})^{-1/4}\Big|_{\phi=\pi}
    =(x_1-x_{1\uparrow}^{-})^{-1/4}e^{-i\pi/4}.
\end{equation}
Hence, the continuation of $F_2(z)$ along the upper contour $\mathcal{C}_1^{+}$ yields
\begin{equation}
    F^{+}_1(x) = C_1\sqrt{\frac{\mathcal{A}(x)}{p(x)}}
    \exp{\left(-\frac{i}{\hbar}\int^x_{x_1} p(\xi)\mathrm{d}\xi\right)}, 
\end{equation}
with the connection coefficient
\begin{equation}\label{eq-C1C2}
  C_1 = C_2 e^{-i\pi/4} e^{\gamma}.
\end{equation}

Similarly, analytic continuation along the lower contour $\mathcal{C}^{-}_1$ depicted in Fig.~\ref{fig8}(b), corresponding to $\phi\rightarrow-\pi$ at $x^{-}_{1\downarrow}$, gives 
\begin{equation}
    F^{-}_1(x)=C^{\prime}_1\sqrt{\frac{\mathcal{A}(x)}{p(x)}}
    \exp{\left(+\frac{i}{\hbar}\int^x_{x_1} p(\xi)\mathrm{d}\xi\right)} 
\end{equation}
with
\begin{equation}\label{eq-C1'C2}
    C^{\prime}_1 = C_2e^{i\pi/4}e^{\gamma}.
\end{equation}
The upper component in the classically allowed region I is therefore given by the superposition of the two quasi-classical branches~\cite{Landau1958}
\begin{equation} \label{eq-F1fromx2}
    \begin{split}
        F_1(x)=&\ F^{+}_1(x)+F^{-}_1(x) \\
        =&\ C_1\sqrt{\frac{\mathcal{A}(x)}{p(x)}}\exp{\left(-\frac{i}{\hbar}\int^x_{x_1} p(\xi)\mathrm{d}\xi\right)} 
            + C^{\prime}_1\sqrt{\frac{\mathcal{A}(x)}{p(x)}}
                \exp{\left(\frac{i}{\hbar}\int^x_{x_1} p(\xi)\mathrm{d}\xi\right)}.    
    \end{split}
\end{equation}
The second term represents the incident wave propagating toward the barrier, while the first term corresponds to the reflected wave propagating away from it. By combining expressions~\eqref{eq-F3},~\eqref{eq-F2fromx3up}, and~\eqref{eq-F1fromx2} for the upper component in different regions with the associated connection formulas~\eqref{eq-C2C3}, \eqref{eq-C1C2}, and \eqref{eq-C1'C2}, we obtain the complete expressions for the upper component in all three regions, as explicitly presented in Eq.~\eqref{eq-F123} of the main text.

To derive the normalization condition~\eqref{eq-Nmod} in the main text, we set the following
\begin{equation}
    C_2 = \frac{\sqrt{\mathcal{N}}}{2}.
\end{equation}
In the nonrelativistic case, the WKB wave function in the classically forbidden region II ($r_2 < r < r_3$) takes the form
\begin{equation}
    u_2(r) = \frac{C_2}{\sqrt{|k(r)|}}\exp{\left(-\int_{r_2}^{r}|k(\xi)|\dd \xi\right)},
\end{equation}
where the local wave number $k(r)$, defined in Eq.~\eqref{eq-kr}, is purely imaginary in this region.
Applying the method detailed in this appendix yields the explicit form of the wave function $u_1(r)$ in the classically allowed region I ($r_1<r<r_2$)
\begin{align}\label{eq-u1}
     u_1(r) &= \frac{C_2e^{-i\pi/4}}{\sqrt{k(r)}}\exp{\left(-i\int_{r_2}^{r}k(\xi)\dd \xi\right)}+\frac{C_2e^{i\pi/4}}{\sqrt{k(r)}}\exp{\left(i\int_{r_2}^{r}k(\xi)\dd \xi\right)}\notag\\
     &=\sqrt{\frac{\mathcal{N}}{k(r)}}\cos\left(\int_{r}^{r_2}k(\xi)\dd \xi-\frac{\pi}{4}\right).
\end{align}
In proton radioactivity, the emitted proton is assumed to be entirely localized within this region before emission, leading to the normalization condition
\begin{equation}\label{eq-nonrel-nom}
    \int_{r_1}^{r_2} |u_1(r)|^2 \mathrm{d}r = 1.
\end{equation}
Substituting Eq.~\eqref{eq-u1} into Eq.~\eqref{eq-nonrel-nom} directly yields the normalization condition presented in the main text.

\section{Relativistic tunneling through a rectangular barrier}\label{app-3}
We consider here an exactly solvable rectangular scalar-vector barrier within the same $1+1$ dimensional Dirac framework. Starting from the Dirac equation~\eqref{eq-Dirac1+1}, we take the scalar and vector potentials to be
\begin{equation}
	V(x) = 
	\begin{cases}
		V_0, & 0 < x < a, \\[12pt]
		0,   & \text{otherwise},
	\end{cases}
	\qquad
	S(x) = 
	\begin{cases}
		S_0, & 0 < x < a, \\[12pt]
		0,   & \text{otherwise},
	\end{cases}
\end{equation}
where $V_0 > 0$ and $S_0 < 0$ are constants. Given our primary focus on barrier penetration, we impose the condition
\begin{equation}
    \epsilon=E-m<S_0+V_0.
\end{equation}

In each region of the rectangular barrier, the scalar and vector potentials are constants. The Dirac equation for the upper and lower components,
\begin{subequations}\label{eq-C-dFdG}
    \begin{align}
        i\hbar \frac{\dd F}{\dd x} =&\ (V-E - m-S)G, \label{eq-C-dFdG-F} \\
        i\hbar \frac{\dd G}{\dd x} =&\ (V-E + m +S)F, 
    \end{align}
\end{subequations}
can therefore be reduced to the second-order differential equations
\begin{subequations} \label{eq-C-d2Fd2G}
    \begin{align}
        \frac{\dd^2 F}{\dd x^{2}} =&\ -\frac{(V-E)^{2} - (m+S)^2}{\hbar^2}F,  \\
        \frac{\dd^2 G}{\dd x^{2}} =&\ -\frac{(V-E)^{2} - (m+S)^2}{\hbar^2}G. 
    \end{align}
\end{subequations}
In the free regions $x<0$ or $x>a$, Eq.~\eqref{eq-C-d2Fd2G} reduces to
\begin{subequations}
    \begin{align}
        \frac{\dd^2F}{\dd x^{2}} =&\ -\frac{E^{2} - m^2}{\hbar^2}F,  \\
        \frac{\dd^2G}{\dd x^{2}} =&\ -\frac{E^2-m^2}{\hbar^2}G. 
    \end{align}
\end{subequations}
The solutions are plane waves with wave number 
\begin{equation}
    k\equiv \frac{\sqrt{E^2-m^2}}{\hbar}.
\end{equation}
For $x<0$, the wave function contains an incident wave and a reflected wave, whereas for $x>a$ only the transmitted wave survives. The upper component can thus be written as
\begin{equation}
	F(x)=
        \begin{cases}
		  e^{ikx}+Re^{-ikx}, & x<0,\\[12pt]
		  Se^{ikx}, & x>a,
		\end{cases}
\end{equation}
where $R$ and $S$ are the reflection and transmission amplitudes, respectively. From Eq.~\eqref{eq-C-dFdG-F}, the corresponding lower component is given by
\begin{equation}
    G(x)=
        \begin{cases}
	       	\frac{\hbar k}{E+m}(e^{ikx}-Re^{-ikx}), & x<0,\\[12pt]
		  \frac{\hbar k}{E+m}Se^{ikx}, & x>a.
		\end{cases}
\end{equation}
Inside the barrier $0<x<a$, the second-order equations become
\begin{subequations}
    \begin{align}
        \frac{{\dd}^{2}F}{\dd x^{2}} =&\ -\frac{(V_0-E)^{2} - (m+S_0)^2}{\hbar^2}F,  \\
        \frac{\dd^2G}{\dd x^{2}} =&\ -\frac{(V_0-E)^{2} - (m+S_0)^2}{\hbar^2}G,
    \end{align}
\end{subequations}
yielding exponential solutions characterized by
\begin{equation}
    \kappa\equiv \frac{\sqrt{(m+S_0)^2 - (V_0-E)^{2}}}{\hbar} = \frac{\sqrt{2m|E_\text{eff}-U_{\text{eff}}|}}{\hbar}.    
\end{equation} 
Here, $E_{\text{eff}}$ and $U_{\text{eff}}$ are defined identically to Eq.~\eqref{eq-eff-def}, but with the spatially dependent potentials $S(x)$ and $V(x)$ replaced by the constants $S_0$ and $V_0$, respectively.
The upper component is then written as
\begin{equation}
    F(x)=Ae^{\kappa x}+Be^{-\kappa x},
\end{equation}
while the lower component follows from Eq.~\eqref{eq-C-dFdG-F} as
\begin{equation}
    G(x)=\frac{-i\hbar \kappa}{E-V_{0}+m+S_0}(Ae^{\kappa x}-Be^{-\kappa x}),
\end{equation}
where $A$ and $B$ are constants to be determined by the boundary conditions.

The matching conditions at the discontinuous boundaries follow directly from the Dirac equation itself. Integrating Eq.~\eqref{eq-Dirac1+1} across a small neighborhood $[x_b-\epsilon,\,x_b+\epsilon]$ of a boundary point $x_b$ (here $x_b=0$ or $a$), where $\epsilon$ is a small positive quantity, one finds
\begin{equation}
    -i\hbar \alpha \left[ \Psi(x_b+\epsilon) - \Psi(x_b-\epsilon) \right] 
    + \int_{x_b - \epsilon}^{x_b + \epsilon} \left[ \beta(m + S) + V - E \right] \Psi \, \dd x = 0.
\end{equation}
If $S(x)$ and $V(x)$ have only finite discontinuities and contain no $\delta$-function singularities, the integral term vanishes in the limit $\epsilon\to0$. The Dirac spinor must therefore be continuous across the boundary,
\begin{equation}
   \lim_{\epsilon \to 0^+} \Psi(x_b - \epsilon) =\Psi(x_b^{-})  = \lim_{\epsilon \to 0^+} \Psi(x_b + \epsilon) = \Psi(x_b^{+}).
\end{equation}
This is different from the Schr\"odinger case, where the matching is usually imposed on the wave function and its derivative. Here the first-order nature of the Dirac equation requires the continuity of the spinor itself.

The continuity condition at $x=0$ leads to
\begin{subequations}
    \begin{align}
	   1+R=&\ A+B,\\
        \frac{\hbar k}{E+m}(1-R)=&\ \frac{-i\hbar \kappa}{E-V_{0}+m+S_0}(A-B).
    \end{align}
\end{subequations}
It is convenient to introduce the dimensionless quantity
\begin{equation}
    u \equiv \frac{E-V_{0}+m+S_0}{E+m},
\end{equation} 
which originates from the spinor matching at the discontinuous boundaries. The coefficients $A$ and $B$ can then be expressed in terms of the reflection amplitude $R$ as
\begin{subequations} \label{eq-ABR}
    \begin{align}
	    A=&\ \frac{1}{2}\left[\left(1+u\frac{ik}{\kappa}\right)+R\left(1-u\frac{ik}{\kappa}\right)\right],\\
        B=&\ \frac{1}{2}\left[\left(1-u\frac{ik}{\kappa}\right)+R\left(1+u\frac{ik}{\kappa}\right)\right].
    \end{align}
\end{subequations}
Similarly, continuity at $x=a$ gives
\begin{subequations}
    \begin{align}
	   Ae^{\kappa a}+Be^{-\kappa a}=&\ Se^{ika},\\
       \frac{-i\hbar \kappa}{E-V_{0}+m+S_0}\left(Ae^{\kappa a}-Be^{-\kappa a}\right)
            =&\ \frac{\hbar k}{E+m}Se^{ika}.
\end{align}
\end{subequations}
Thus, $A$ and $B$ can also be written in terms of the transmission amplitude $S$:
\begin{subequations}\label{eq-ABS}
    \begin{align}
	   A=&\ \frac{S}{2}\left(1+u\frac{ik}{\kappa}\right)e^{ika-\kappa a},\\
       B=&\ \frac{S}{2}\left(1-u\frac{ik}{\kappa}\right)e^{ika+\kappa a}.
    \end{align}
\end{subequations}
Eliminating $A$ and $B$ between Eqs.~\eqref{eq-ABR} and~\eqref{eq-ABS}, one obtains
\begin{subequations}
    \begin{align}
	    \left(1+u\frac{ik}{\kappa}\right)+R\left(1-u\frac{ik}{\kappa}\right)
            =&\ S\left(1+u\frac{ik}{\kappa}\right)e^{ika-\kappa a},\\
        \left(1-u\frac{ik}{\kappa}\right)+R\left(1+u\frac{ik}{\kappa}\right)
            =&\ S\left(1-u\frac{ik}{\kappa}\right)e^{ika+\kappa a}.
    \end{align}
\end{subequations}
Solving this system for the transmission amplitude $S$ yields
\begin{equation}
        S=\frac{e^{-ika}\left[\left(1+u\frac{ik}{\kappa}\right)^2-\left(1-u\frac{ik}{\kappa}\right)^2\right]}
            {\left(1+u\frac{ik}{\kappa}\right)^2 e^{-\kappa a}-\left(1-u\frac{ik}{\kappa}\right)^2 e^{\kappa a}}
        =\frac{e^{-ika}2u\frac{k}{\kappa}}{2u\frac{k}{\kappa}\cosh{\kappa a}+i[1-u^2\frac{k^2}{\kappa^2}]\sinh{\kappa a}}.
\end{equation}

From the definition of the probability current in Eq.~\eqref{eq-jDirac}, the transmitted and incident currents are given by
\begin{subequations}
\begin{align}
    j_3=&\ \frac{2\hbar k}{E+m}|S|^2, \\
    j_1=&\ \frac{2\hbar k}{E+m}.
\end{align}
\end{subequations}
Therefore, the transmission probability $T$ is
\begin{equation} \label{eq-T-rec}
    \begin{split}
        T=\frac{j_3}{j_1}=|S|^{2}
        =&\ \frac{4k^{2}\kappa^{2}u^{2}}{(k^{2}u^{2}-\kappa^{2})^{2}\sinh^{2}\kappa a+4k^{2}u^2\kappa^{2}\cosh^{2}\kappa a} \\
        =&\ \frac{4k^{2}\kappa^{2}u^2}{(k^{2}u^{2}+\kappa^{2})^{2}\sinh^{2}\kappa a+4k^{2}\kappa^{2}u^{2}} \\
        =&\ \frac{1}{1+\frac{(k^2u^2+\kappa^2)^2}{4k^2\kappa^2u^2}\sinh^2\kappa a},
    \end{split}
\end{equation}
where the identity $\cosh^2x-\sinh^2x=1$ has been used.

This exact result provides a useful complement to the WKB analysis for a continuous barrier. In the latter case, the relativistic dynamical effect is mainly reflected in the effective potential $U_{\text{eff}}$ entering the local momentum. For the discontinuous rectangular barrier considered here, however, it appears not only in the local momenta $k$ and $\kappa$, where the latter explicitly involves $U_{\text{eff}}$, but also in the boundary matching of the Dirac spinor through the factor $u$. The rectangular-barrier solution thus suggests that the relativistic dynamical effect may be viewed as arising from two closely related ingredients: the relativistic dispersion relation, which determines the local momenta, and the spinor structure of the Dirac wave function, which enters through the boundary matching conditions.

Furthermore, for a wide barrier ($\kappa a \gg 1$), since $\sinh(\kappa a) \approx \frac{1}{2}e^{\kappa a} \gg 1$, the transmission probability simplifies to 
\begin{equation}
    T \approx \frac{16k^2\kappa^2u^2}{(k^2u^2+\kappa^2)^2} e^{-2\kappa a}.
\end{equation}
This exponentially suppressed behavior resembles the standard WKB tunneling result in Eq.~\eqref{eq-TDirac}.

Additionally, in the nonrelativistic limit $|V_0|, |S_0|\ll m$, the spinor matching factor approaches unity ($u\to1$). Consequently, Eq.~\eqref{eq-T-rec} reduces to the standard nonrelativistic form
\begin{equation}
    T=\frac{1}{1+\frac{(k^2+\kappa^2)^2}{4k^2\kappa^2}\sinh^2\kappa a},
\end{equation}
where the remaining difference lies solely in the definitions of $k$ and $\kappa$, with $\kappa$ now incorporating $U_{\text{eff}}$.

\section{Bohr-Sommerfeld quantization condition}\label{app-4}

We now consider the bound-state problem in such an effective potential well $U_{\text{eff}}(x)$, as shown in Fig~\ref{fig10}. In the classical forbidden region III where the momentum $p(x)$ is purely imaginary, only an exponentially decaying wave exists and thus the upper component $F(x)$ is written as 
\begin{equation}
        F_3(x) = C_3 \sqrt{\frac{\mathcal{A}(x)}{|p(x)|}} 
             \exp\!\left( -\frac{1}{\hbar} \int_{x_2}^{x} |p(\xi)| \, \mathrm{d}\xi \right),\quad x_2<x.
\end{equation}
Following the same method in Appendix~\ref{app-2} to continue in the complex plane across the turning point $x_2$, it yields the oscillatory wave in region II
\begin{equation}
     F_2^{(1)}(x)=2C_2 \sqrt{\frac{\mathcal{A}(x)}{p(x)}}\cos\left(\frac{1}{\hbar}\int^{x_2}_{x} p(\xi) \mathrm{d}\xi -\frac{\pi}{4}\right),\quad x_1<x<x_2,
\end{equation}
with the coefficient relation
\begin{equation}
    C_2=C_3.
\end{equation}

\begin{figure}
    \centering
	\includegraphics[width=0.40\linewidth]{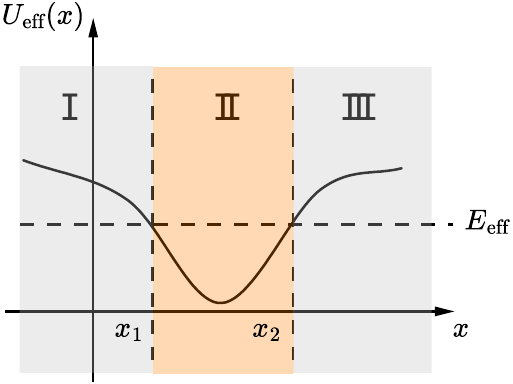}
    \caption{Effective potential well $U_{\mathrm{eff}}(x)$ with turning points $x_1$ and $x_2$ defined by $E_{\mathrm{eff}}=U_{\mathrm{eff}}(x)$. The grey-shaded regions I and III are classically forbidden, while the orange-shaded region II is classically allowed.}
    \label{fig10}
\end{figure} 

Since only an exponentially decaying wave exists in the classically forbidden region I, the upper component $F(x)$ takes the form
\begin{equation}
           F_1(x) = C_1 \sqrt{\frac{\mathcal{A}(x)}{|p(x)|}} 
             \exp\!\left( -\frac{1}{\hbar} \int_{x}^{x_1} |p(\xi)| \, \mathrm{d}\xi \right),\quad x<x_1.
\end{equation}
Through the continuation in the complex plane to bypass the turning point $x_1$, we obtain a similar oscillatory wave with a different integration limit $x_1$ in region II
\begin{equation}
           F^{(2)}_2(x) = 2C'_2 \sqrt{\frac{\mathcal{A}(x)}{p(x)}}\cos\left(\frac{1}{\hbar}\int^{x}_{x_1} p(\xi) \mathrm{d}\xi -\frac{\pi}{4}\right),\quad x_1<x<x_2,
\end{equation}
with the coefficient relation
\begin{equation}
    C'_2=C_1.
\end{equation}

Since both $F^{(1)}_2(x)$ and $F^{(2)}_2(x)$ correspond to the upper component $F(x)$ in region II, their consistency requires that the sum of their phases must be an integer multiple of $\pi$, which yields the Bohr-Sommerfeld quantization condition
\begin{equation}\label{quantization condition}
    \frac{1}{\hbar} \int_{x_1}^{x_2} p(x)  \dd x=(n+\frac{1}{2})\pi,\quad n=0,1,2\cdots
\end{equation}
with coefficient relation 
\begin{equation}
    C_2=C_2^{\prime}(-1)^n.
\end{equation} 

In comparison to the nonrelativistic case, the Bohr-Sommerfeld quantization condition~\eqref{quantization condition} in the relativistic framework takes the same form but incorporates a modified definition of the local momentum $p(x)$. In addition, the relativistic dynamical effect is also mainly reflected in the effective potential $U_{\text{eff}}$ entering the local momentum.

\end{widetext}

\bibliography{refe.bib}

\end{document}